\documentclass[twocolumn,preprintnumbers,amsmath,amssymb,superscriptaddress]{revtex4}
\usepackage{bm}%
\usepackage[dvips]{graphicx, color}

\usepackage{comment}

\expandafter\ifx\csname package@font\endcsname\relax\else
 \expandafter\expandafter
 \expandafter\usepackage
 \expandafter\expandafter
 \expandafter{\csname package@font\endcsname}%
\fi

\begin{document}
\title{Ferromagnetism of cobalt-doped anatase TiO$_2$ studied by bulk- and surface-sensitive soft x-ray magnetic circular dichroism}%

\author{V.~R.~Singh}
\email{vijayraj@wyvern.phys.s.u-tokyo.ac.jp}
\affiliation{Department of Physics, University of Tokyo, Bunkyo-ku, Tokyo 113-0033, Japan}

\author{K.~Ishigami}
\affiliation{Department of Physics, University of Tokyo, Bunkyo-ku, Tokyo 113-0033, Japan}

\author{V.~K.~Verma}
\affiliation{Department of Physics, University of Tokyo, Bunkyo-ku, Tokyo 113-0033, Japan}

\author{G.~Shibata}
\affiliation{Department of Physics, University of Tokyo, Bunkyo-ku, Tokyo 113-0033, Japan}

\author{Y.~Yamazaki}
\affiliation{Department of Physics, University of Tokyo, Bunkyo-ku, Tokyo 113-0033, Japan}

\author{T.~Kataoka}
\affiliation{Department of Physics, University of Tokyo, Bunkyo-ku, Tokyo 113-0033, Japan}

\author{A.~Fujimori}
\affiliation{Department of Physics, University of Tokyo, Bunkyo-ku, Tokyo 113-0033, Japan}

\author{F.-H.~Chang}
\affiliation{National Synchrotron Radiation Research Center (NSRRC), Hsinchu 30076, Taiwan, Republic of China}

\author{D.-J.~Huang}
\affiliation{National Synchrotron Radiation Research Center (NSRRC), Hsinchu 30076, Taiwan, Republic of China}

\author{H.-J.~Lin}
\affiliation{National Synchrotron Radiation Research Center (NSRRC), Hsinchu 30076, Taiwan, Republic of China}

\author{C.~T.~Chen}
\affiliation{National Synchrotron Radiation Research Center (NSRRC), Hsinchu 30076, Taiwan, Republic of China}

\author{Y.~Yamada}
\affiliation{Institute for Materials Research, Tohoku University, Sendai 980-8577, Japan}

\author{T.~Fukumura}
\affiliation{Department of Chemistry, University of Tokyo, Bunkyo-ku, Tokyo 113-0033, Japan}
\affiliation{PRESTO, Japan Science and Technology Agency, Kawaguchi 332-0012, Japan}

\author{M.~Kawasaki}
\affiliation{Institute for Materials Research, Tohoku University, Sendai 980-8577, Japan}
\affiliation{WPI-AIM Research, Tohoku University, Sendai 980-8577, Japan}
\affiliation{Quantum-Phase Electronics Center and Department of Applied Physics, University of Tokyo, Tokyo 113-8656, Japan}
\affiliation{CREST, Japan Science and Technology Agency, Tokyo 102-0075, Japan}

\date{\today}

\begin{abstract}
 We have studied magnetism in anatase Ti$_{1-x}$Co$_x$O$_{2-\delta}$ ({\it x} = 0.05) thin films with various electron carrier densities, by soft x-ray magnetic circular dichroism (XMCD) measurements at the Co $L_{2,3}$ absorption edges. For electrically conducting samples, the magnetic moment estimated by XMCD was  $<$ 0.3 $\mu_B$/Co using the surface-sensitive total electron yield (TEY) mode, while it was 0.3-2.4 $\mu_B$/Co using the bulk-sensitive total fluorescence yield (TFY) mode. The latter value is in the same range as the saturation magnetization 0.6-2.1 $\mu_B$/Co deduced by SQUID measurement. The magnetization and the XMCD intensity increased with carrier density, consistent with the carrier-induced origin of the ferromagnetism.

\end{abstract}


\maketitle

Semiconductors partially substituted with magnetic ions are called diluted magnetic semiconductors (DMSs) and are expected to be useful in spintronics devices, where electron spins can be controlled by electric field and/or by photons. Ferromagnetic DMS's with Curie temperatures ($T_C$'s) higher than room temperature are highly desirable for the development of spintronic devices. To date, much work in this area has been done, mainly on II-VI and III-V compounds doped with magnetic ions such as (Cd,Mn)Te \cite{1} and (Ga,Mn)As \cite{2,3}, but their $T_C$'s are far below room temperature. Ferromagnetism was observed in Mn-based zinc-blende II-VI compounds such as (Cd,Mn)Te after the result of carrier induced ferromagnetism \cite{90}. Kuroda {\it et al.}\cite{50,95} reported that Cr rich phases of (Zn,Cr)Te showed room temperature ferromagnetism, causing a stimulation of II-VI DMS. Ferromagnetism was also observed in (Ga,Mn)As \cite{2,3}. It was theoretically suggested \cite{40} that the co-doping of magnetic semiconductors with shallow impurities affects the self-assembly of magnetic nanocrystals during epitaxy, and therefore modifies both the global and local magnetic behavior of the material.  This concept was also qualitatively corroborated by experimental data for (Cd, Mn, Cr)Te \cite{45} and (Ga, Mg, Fe)N \cite{60, 70, 80}. However, origin of ferromagnetism at room-temperature is controversial so far.  Recently, Matsumoto {\it et al.}\cite {100,105} reported the occurrence of room temperature ferromagnetism in Co-doped anatase TiO$_2$ films. According to Fukumura {\it et al.} \cite{5,7} the high electron carrier densities and Co content favor the ferromagnetic phase in Co-doped rutile TiO$_2$ at 300 K. Room-temperature ferromagnetism was also reported in such materials as (Ga,Mn)N \cite{22} and (Al,Cr)N \cite{23}. The near edge x-ray absorption fine structure study of Co-doped TiO$_2$ by Griffin {\it et al.} \cite{24} claims that ferromagnetism is due to $d$-$d$ double exchange mediated by tunneling of $d$ electrons within the impurity band. Some studies that also claim the ferromagnetism of Co-doped TiO$_2$ is due to Co metal clusters \cite{25,26,27,28}. The recent theoretical study by Calderon {\it et al.} \cite{29}, electric field-induced anomalous Hall effect (AHE) study by Yamada {\it et al.} \cite{15} and x-ray photoemission spectroscopy study by Ohtsuki {\it et al.} \cite{30} suggested the ferromagnetism of Co-doped TiO$_2$ is due to carrier mediated. However, direct information about the magnetization as a function of carrier density has been lacking. Soft x-ray magnetic circular dichroism (XMCD) at the Co 2$p\rightarrow3d$ absorption (Co $L_{2,3}$)  edge is a powerful technique to clarify this issue because it is an element-specific magnetic probe \cite{28}. A previous XMCD study on rutile Co-doped TiO$_2$ by Mamiya {\it et al.} has revealed that the ferromagnetism is not due to segregated Co metal clusters but is due to Co$^{2+}$ ions in the TiO$_2$ matrix \cite{6}. However, the XMCD signal intensities were an order of magnitude lower than that expected from the bulk magnetization \cite{6}. In a more recent work \cite{7}, we performed x-ray absorption spectroscopy (XAS) and XMCD studies on rutile Co-doped TiO$_2$ not only by the surface-sensitive total electron yield (TEY) mode but also the bulk-sensitive total fluorescence yield (TFY) mode and found that Co ions in the bulk indeed have a large magnetic moment of 0.8-2.2 $\mu_B$/Co.

In this work we have extended the same approach to anatase Co-doped TiO$_2$ and studied correlation between magnetism and transport properties. Magnetization measurements of anatase Ti$_{1-x}$Co$_x$O$_{2-\delta}$ thin films reveal ferromagnetic hysteresis behavior in the \textsl{M-H} loop at room temperature with a saturation magnetization. In the bulk region probed by the TFY mode, strong XMCD spectra with similar spectral line shapes were obtained for all the samples. The magnetization and the XMCD intensity increased with carrier density, consistent with the carrier-induced origin of the ferromagnetism.

 Anatase Ti$_{1-x}$Co$_x$O$_{2-\delta}$ epitaxial thin films with $x$ = 0.05 were synthesized by the pulsed laser deposition method on LaAlO$_3$ (001) substrates at 523 K and oxygen pressures ($P_{{\rm O}_2}$) of 5 $\times$ $10^{-7}$, 1 $\times$ $10^{-6}$ and 2 $\times$ $10^{-6}$ Torr. The resistivity increases in this order and these samples are hereafter referred to metallic, intermediate, insulating samples, respectively. The carrier densities $n_e$ were 4.1 $\times$ $10^{19}$, 1.1 $\times$ $10^{19}$  and 4.0 $\times$ $10^{18}$ cm$^{-3}$, respectively. Segregation of secondary phases were not observed under careful inspections by x-ray diffraction (XRD) and transmission electron microscopy (TEM) \cite{15} of $\sim$40 nm thick films \cite{15}. Reflection high-energy electron diffraction was monitored during the {\it in-situ} growth. An intensity oscillation was observed at the initial stage of the growth. We have confirmed that Co distribution along the film thickness direction in our films is uniform using TEM \cite{15}, unlike the inhomogeneous distribution in films prepared on Si demonstrated using atom probe tomography by Larde {\it et al} \cite {31}. Ferromagnetism at room temperature was confirmed by Hall-effect measurements and magnetization measurements. XAS and XMCD measurements were performed at the BL-11A  beamline of the National Synchrotron Radiation Research Center, Taiwan. In XMCD measurements, magnetic fields (H) were applied  parallel to the direction of anatase (001). XAS and XMCD spectra were obtained in the TEY and TFY modes and probing depths were $\sim$5 and 100 nm, respectively.

\begin{figure}[htbp]
\begin{center}
\includegraphics[width=08cm]{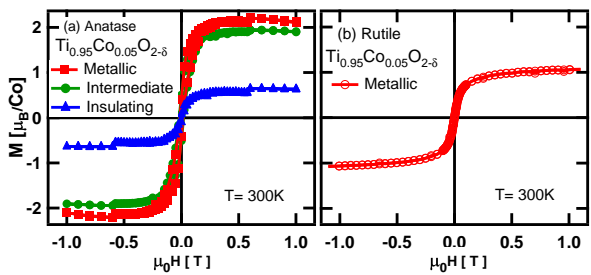}
\caption{(Color online) \textsl{M-H} curves of Ti$_{0.95}$Co$_{0.05}$O$_{2-\delta}$ at 300 K. (a) Metallic, intermediate and insulating anatase samples. (b) Metallic rutile sample.}
\end{center}
\end{figure}
Figure 1(a) shows the magnetization curves of anatase Ti$_{1-x}$Co$_x$O$_{2-\delta}$ ($x$ = 0.05) at 300 K for various carrier densities ($n_e$). The $n_e$ for metallic, intermediate and insulating samples were 4.1 $\times$ $10^{19}$, 1.1 $\times$ $10^{19}$ and 4.0 $\times$ $10^{18}$ cm$^{-3}$, respectively. That of metallic rutile thin films which has the carrier density of 7 $\times$ $10^{21}$ cm$^{-3}$ is also shown in Fig 1(b). The saturation magnetization of the anatase sample is 0.6-2.1 $\mu_B$/Co with a coercive force of $\sim$100 to 200 Oe. In the \textsl{M(H)} measurements, magnetic field was applied parallel to the the direction of anatase (001). Anomalous Hall-effect (AHE) measurements for anatase Ti$_{1-x}$Co$_x$O$_{2-\delta}$ with various $n_e$ also show similar magnetic field dependences \cite{15}. From Fig. 1, it is clear that the magnetization of the anatase thin films is larger than the rutile thin films, which may be attributed to the fact that anatase films in this study have a mobility $\sim$2-11 cm$^2$V$^{-1}$s$^{-1}$ which is two orders of the magnitude higher than the mobility of rutile thin films \cite{5}.

\begin{figure}[htbp]
\begin{center}
\includegraphics[width=08cm]{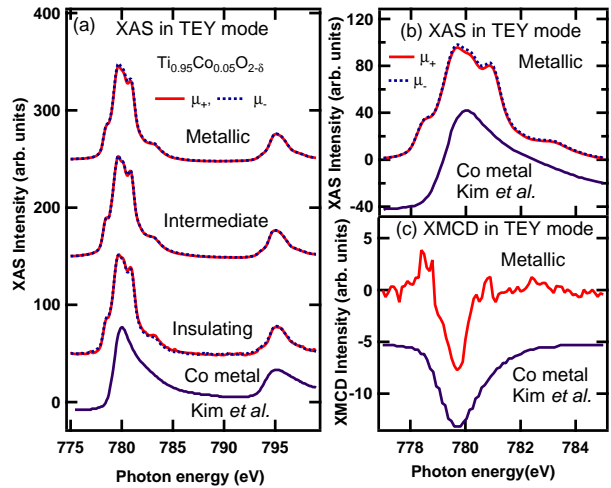}
\caption{(Color online) Co $L_{2,3}$-edge of anatase Ti$_{0.95}$Co$_{0.05}$O$_{2-\delta}$ taken in the TEY mode at \textsl {T} = 300 K and \textsl {H} = 1 T. (a) XAS. (b),(c) XAS and XMCD spectra of the metallic anatase Ti$_{0.95}$Co$_{0.05}$O$_{2-\delta}$ sample. The XAS and XMCD spectra of Co metal by Kim {\it et al.}\cite{28} are shown for comparison.}
\end{center}
\end{figure}

In Fig. 2(a), we show the Co $L_{2,3}$-edge XAS (metallic, intermediate and insulating thin films) and Fig. 2(b)-(c) XAS and XMCD spectra of (metallic thin film) anatase Ti$_{1-x}$Co$_x$O$_{2-\delta}$ obtained in the TEY mode. In the figure, $\mu_+$ and $\mu_-$ refer to the absorption coefficients for photon helicity parallel and antiparallel to the Co majority spin direction, respectively. The XMCD spectra $\Delta\mu$ = $\mu_+$ - $\mu_-$ have been corrected for the degree of circular polarization. The XAS and XMCD spectra of the metallic anatase Ti$_{1-x}$Co$_x$O$_{2-\delta}$ sample showed multiplet features as shown by Fig 2(a)-(c), which is similar to Mamiya {\it et al.} \cite{6} and agree with our $D_{2h}$ high-spin crystal-field symmetry cluster model calculations using the parameter values : Charge-transfer energy ($\Delta$)= 4 eV, On-site 3$d$-3$d$ Coulomb energy ($U_{dd}$)=5 eV, 3$d$-2$p$ Coulomb energy ($U_{dc}$)= 7 eV, Hopping integral between the Co 3$d$ and O 2$p$ orbitals of E$_g$ symmetry ($V_{{\rm E}_g}$)= 1.1 eV and  Crystal-field splitting (10{\it Dq})= 0.9 eV.
 The multiplet features of the XMCD spectra show almost one-to-one correspondence to those in the XAS spectra. The spectral line shapes of the XAS and XMCD spectra for the metallic and intermediate anatase Ti$_{1-x}$Co$_x$O$_{2-\delta}$ samples are also similar to those of rutile Co-doped  TiO$_2$ results which were reported in our previous work \cite{6,7}. For the insulating sample, we observed an XAS spectrum similar to those of the metallic and intermediate samples. The estimated magnetic moments for all samples obtained from XMCD in the TEY mode were $<$ 0.3 $\mu_B$/Co. These values are larger than the 0.1 $\mu_B$/Co which is reported by Mamiya {\it et al.} \cite{6}, but they are still smaller than the saturation magnetic moments 0.6-2.1 $\mu_B$/Co deduced from magnetization measurements. The XAS and XMCD spectra of Co metal is also shown at the bottom of Fig. 2 (a)-(c) for comparison. It is demonstrated that the present XAS and XMCD spectra of Co-doped TiO$_2$ are distinctly different from Co metal.

\begin{figure}[htbp]
\begin{center}
\includegraphics[width=08cm]{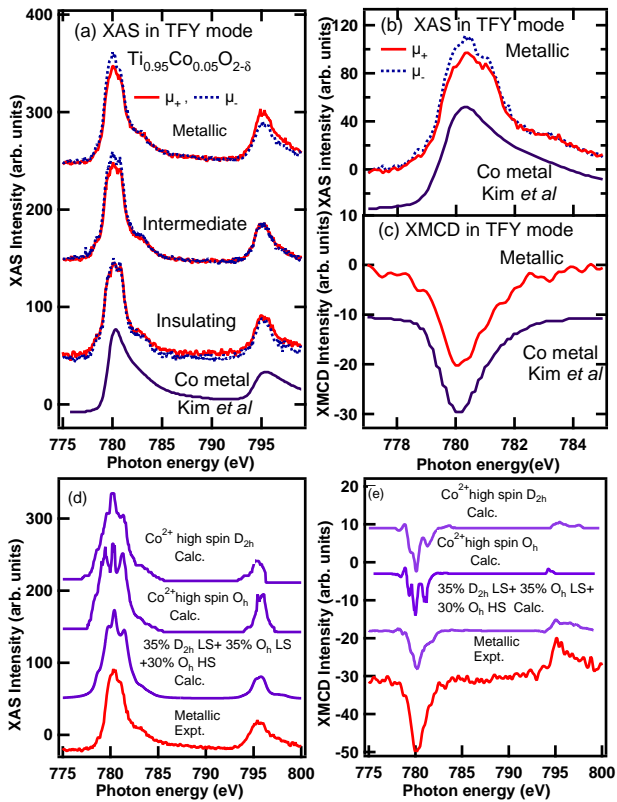}
\caption{(Color online) Co $L_{2,3}$-edge of anatase Ti$_{0.95}$Co$_{0.05}$O$_{2-\delta}$ taken in the TFY mode at \textsl {T} = 300 K and \textsl {H} = 1 T. (a) XAS. (b),(c) XAS and XMCD spectra of anatase Ti$_{0.95}$Co$_{0.05}$O$_{2-\delta}$ for  metallic sample in the TFY mode. The XAS and XMCD spectra of Co metal by Kim {\it et al.} \cite{28} are shown for comparison. (d),(e) Comparison of XAS and XMCD spectra shown in (b) and (c) with cluster-model calculation \cite{34}.}
\end{center}
\end{figure}

\begin{table}[ht]
\caption{Electronic structure parameters for anatase Co-doped TiO$_2$ thin film used in the cluster-model calculations in units of eV to analyze.}
\centering 
\begin{tabular}{c c c c c c c c} 
\hline 
Crystal-field symmetry & Spin & $\Delta$ & $U_{dd}$ & $U_{dc}$ & $V_{{\rm E}_g}$  & 10{\it Dq} & Weight{(\%)}\\ [0.5ex] 
\hline 
$D_{2h}$ & Low  & 4 & 5 & 7 & 1.1 & 1.1$-$1.2 & 35 \\ 
$O_h$ & Low  & 3 & 6 & 7.5 & 1.1 & 1.1$-$1.2 & 35\\ [1ex] 
$O_h$ & High  & 2 & 5 & 7.5 & 1.1& 0.8$-$0.9  & 30\\ [1ex] 
\hline 

\end{tabular}

\label{table:nonlin} 
\end{table}
Figures 3(a),(b) and (c) show the Co $L_{2,3}$ XAS and XMCD spectra of the same samples taken in the TFY mode. From the figure, it is clear that the XMCD intensities are much higher than those taken in the TEY mode. The large difference between the bulk-sensitive TFY mode with $\sim$100 nm probing depth and the surface-sensitive TEY mode with $\sim$5 nm probing depth suggests that there is a magnetically dead layer of $\sim$5 nm thickness or more at the surface of the samples as in the case of  rutile \cite{6,7}. The presence of a surface dead layer of $\sim$5 nm thickness is consistent with the recent measurements of the film-thickness dependence of AHE \cite{32}. The spectral line shapes of the XAS and XMCD spectra of all the samples taken in the TFY mode show broad features with spectral line shapes  similar to those of  rutile Co-doped TiO$_2$ \cite{7}. Both magnetization and XMCD intensity increased with carrier density. This is consistent with spin alignment arises due to the interaction of local spins with the spin polarized free carriers, in which carrier-mediated ferromagnetism and ferromagnetic ordering is realized. Yamada {\it et al.}\cite{15} have also demonstrated electrically induced ferromagnetism at room-temperature in anatase Ti$_{1-x}$Co$_x$O$_{2-\delta}$, by means of electric double layer gating resulting in high density electron accumulation ($>$10$^{14}$ cm$^{-2}$). By applying a gate voltage of a few volts, a low-carrier paramagnetic state was transformed to a high-carrier ferromagnetic state. This also supports theoretically as well as experimentally the idea that the ferromagnetism originates from a carrier-mediated mechanism \cite{15,29}. The broadening of the TFY spectra may be due to the randomly displaced positions of Co atoms, which leads to in various local structures as suggested by the anomalous X-ray scattering study of  Matsumura {\it et al.} \cite{33}. The experimental XAS and XMCD spectra are distinctly different from Co metal and show qualitatively good agreement with the calculated spectra for the Co$^{2+}$ in random crystal fields \cite{34}, where the calculations were done using the various electronic structure parameters as listed in Table I.

\begin{figure}[htbp]
\begin{center}
\includegraphics[width=08cm]{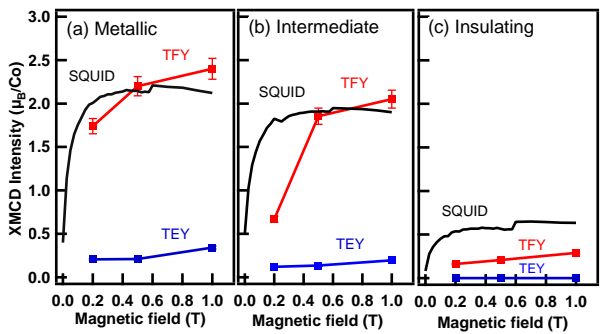}
\caption{(Color online)  Magnetization as a function of magnetic field obtained from the XMCD intensities of anatase Ti$_{0.95}$Co$_{0.05}$O$_{2-\delta}$ compared with \textsl{M-H} curves obtained using a SQUID.}
\end{center}
\end{figure}
Figure 4 shows magnetization versus magnetic field curves estimated from the XMCD spectra obtained in the TEY and TFY modes using sum rules \cite{6}, as compared with the \textsl{M-H} curves measured using a SQUID. We have divided the obtained spin-magnetic moment by a correction factor of 0.92 given by Teramura {\it et al.}\cite{35}.  The Co magnetic moment is found to be obviously much larger in the bulk region than in the surface region. These results are also consistent with the x-ray photoemission spectroscopy study by Yamashita {\it et al.} \cite{14}. Since we know that TFY suffers from self-absorption and therefore it will saturate the XAS signal. This saturated XAS signal will reduce XMCD signal. Because of this very fact, we can conclude that the real value of magnetic moment in bulk should be even higher than the measured TFY value in metallic and intermediate samples which are reported in the present work. Accordingly, our observation by using the TEY and TFY modes are validated. The magnetic moment obtained from cluster-model calculation (Fig.3) is 1.6 $\mu_B$/Co, which is similar to the magnetization of $\sim$2 $\mu_B$/Co deduced from the TFY results and the SQUID measurement. These results suggest that the Co ions in the bulk region are responsible for the ferromagnetism in anatase Ti$_{1-x}$Co$_x$O$_{2-\delta}$.

In conclusion, we have studied the ferromagnetism of cobalt-doped anatase TiO$_2$ thin films using element-specific XMCD at the Co $L_{2,3}$ edges in both the surface-sensitive TEY and bulk-sensitive TFY modes. The large magnetic moment of the Co ions, 0.6-2.4 $\mu_B$/Co, was observed by the TFY method. The carrier-induced origin of ferromagnetism at room-temperature in anatase Ti$_{1-x}$Co$_x$O$_{2-\delta}$ is supported by the XMCD study of the samples with different carrier concentration. According to the spectra taken in the TFY mode, the positions of Co$^{2+}$ atoms seem to be displaced from  the  regular Ti$^{4+}$ sites, resulting in random crystal fields. Good agreement is demonstrated not only in magnetization and AHE but also in the magnetic field dependences of XMCD.  The magnetic moment values deduced with the TEY mode was $<$ 0.3 $\mu_B$/Co, indicating the presence of a magnetically dead layer of $\sim$5 nm thickness at the sample surfaces.

This work was supported by a Grant-in-Aid for Scientific Research in Priority Area ``Creation and Control of Spin Current'' (19048012) from MEXT, Japan, Grant-in-Aid for Scientific Research (S 22224005) from JSPS and TF was supported by the Funding Program for Next Generation World-Leading Researchers.


\begin{thebibliography}{references}
\bibitem{34} The calculated spectra are the weighted sum of 35\% of $D_{2h}$ low spin (LS), 35\% of $O_h$ low spin (LS), 30\% of $O_h$ high spin (HS).
\end{thebibliography}
\end{document}